\documentstyle[psfig,epsf,conf-X,10pt]{article}
\newcommand{\pspc}{{\it PSPC} }
\newcommand{\asca}{{\it ASCA} } 
\newcommand{\rosat}{{\it ROSAT} }
\newcommand{\mnras}{MNRAS }
\newcommand{\etal}{{\it et al.} }
\newcommand{\bsax}{{\it BeppoSAX} }

\begin{document} 
\small
\heading{%
%
What Kind of Obscured AGN Make the CXRB?
}
\par\medskip\noindent
\author{%
Tahir Yaqoob$^{1}$
}
\address{%
Laboratory for High Energy Astrophysics, NASA/GSFC, code 662,
Greenbelt, MD 20771, USA. (Universities Space Research Association.)
}
%

\begin{abstract}
While it is understood that the origin of the Cosmic X-ray
Background (CXRB) can be explained in terms of a combination of
obscured and unobscured AGN, the identity of these obscured AGN
showing up in large numbers at faint fluxes remains unclear.
Deep surveys are showing up increasingly
large numbers of AGN at the faintest fluxes which have {\it broad optical
lines but flat X-ray spectra}. It is commonly assumed that there
must a new population of sources making a 
substantial contribution to the CXRB.
We show that we need not look further
than our own backyard to find familiar examples of such 
sources and there is no compelling reason to invoke a mysterious,
new population.
\end{abstract}
\section{Introduction }

The last decade has seen some major progress towards a solution of the
Cosmic X-ray Background (CXRB) problem. As a
result of  instrumentation
with improved
sensitivity and energy resolution and the realization that a
combination of heavily absorbed and unabsorbed active galaxy populations are
required to make
the CXRB \cite{wolt89}, \cite{mad94}, it is now possible to easily reproduce the shape of the
CXRB, at least in the $\sim 3$-100 keV band 
(e.g. \cite{gill99a}, \cite{coma99} and other papers in these proceedings).
Previously, in the 1980s, spectral
measurements of heavily absorbed AGN spectra were
poor and rare since such spectra were associated with X-ray weak
Seyfert  2 galaxies, so models of the CXRB predicted a spectrum which
was too steep compared to that which was observed.
X-ray absorption provides the flattening of the spectra of some of the
sources necessary to account for the CXRB spectrum.
 More details of the
state of the art of the latest
models can be found within these proceedings and references therein.

However, a few nagging and important problems remain.
The first problem is what to identify this X-ray absorbed population
with. Are they just the regular Seyfert 2 galaxies that we see in our
local Universe? Increasing evidence since the early 1990s has pointed
to a {\it new}, as yet unobserved, population of flat X-ray spectra,
Narrow Emission-Line Galaxies (NELG), dominating the CXRB contribution at
the faintest fluxes (e.g. \cite{laha93}, \cite{miya94}, \cite{pear97}). There has been a massive effort
in the past decade by many different
groups around the world to try and directly resolve the X-ray background
with deep surveys using X-ray imaging telescopes like \rosat and \asca
followed up by optical identification programs, complemented with
fluctuations analyses and cross-correlation studies 
(e.g. \cite{boyl93}, \cite{grif96}, \cite{geor97}, \cite{alma97}, 
\cite{akiy98}, \cite{boyl98},
\cite{cagn98}, \cite{gend98}, \cite{giom98}, \cite{ueda99}, 
\cite{yama99}, \cite{fior99}, and
these proceedings).
The deepest \rosat HRI observations (1.4 Ms) combined with optical
follow-ups indicate that the earlier associations of the faintest
CXRB contributions with NELG may be questionable
 \cite{Has99a}; but see also \cite{mach99}.
About 80\% of the soft X-ray background has been directly
resolved into discrete sources in the \rosat band and
$> 80\%$ of these are claimed to be broad-line AGN from HRI
deep field data \cite{Has99a} but only $31\%$ from the 
deepest \pspc data \cite{mach99}.

In the hard band all except one study resolves $\sim 30\%$ of the
2--10 keV CXRB into a discrete source population with flat X-ray
spectra. The one exception is the \asca fluctuations study by 
\cite{yama99} 
which goes down to a limiting 2--6 keV flux of
$\sim 10^{-14} \rm \ erg \ cm^{-2} \ s^{-1}$ and claims to
resolve $\sim 60\%$ of the CXRB in this energy band. The measured photon
index of the fluctuations is $\Gamma = 1.26 \pm 0.12$. As pointed out by
\cite{yama99} the deeper the flux limit of the hard surveys, the
flatter the X-ray slope of the CXRB contributors.
Optical follow-up studies of all these hard surveys are still in
progress but preliminary findings all come up with the same
result:
that a large percentage of the sources, maybe as much as 50\% turn out
to be {\it broad-line flat X-ray spectra} sources.
This may be surprising to some because broad-line AGN are
traditionally associated with steep X-ray spectra with little
absorption. We shall show in this paper that this view is antiquated
and thus it is premature to conclude from this that there must
be a new, undiscovered population of sources making the CXRB.
We note that at this flux level, all that can be said about the X-ray spectra
is that they are flat; it is possible to model the X-ray
spectra with just a single power-law only, which may in
fact be mimicking a much more complex spectrum.

The second problem concerns the source counts,
$\log{N}/ \log{S}$ (i.e. the differential number of
sources contributing to the CXRB at a given flux), or
equivalently, the X-ray luminosity function (XLF).
If we use the soft X-ray band XLF (0.5--2 keV) measured with, for example,
\rosat and use it to predict the XLF in the hard X-ray band
(2--10 keV) we get a significant discrepancy
compared with observation if we use the same
assumptions about the source spectra which are used to model the
CXRB spectrum. Likewise if we use the hard band XLF to predict the
soft band XLF the result does not agree with observation. This
problem has been noted by many authors 
(e.g. \cite{coma99}, \cite{gill99a}, \cite{giom98}, \cite{wilm99}, 
other papers in these proceedings).
It has even been demonstrated that the discrepancy can be resolved
if we do not insist that the broad-line objects contributing
to the CXRB in the \rosat band have
steep X-ray spectra \cite{giom98} in the hard band. 
Again, due to a misconception
about the X-ray spectra of broad-line AGN, a new source population is
invoked.

In summary, the two major problems are that 
a significant fraction of the flat X-ray
spectrum sources making the CXRB are turning out to be broad-line
AGN rather than narrow-line AGN, and that the soft XLF does not
correctly predict the hard XLF and vice-versa.
In the remainder of this paper we examine the
critical assumptions which go into modeling the CXRB spectrum
and XLF and show that we need look no further than our
own backyard to find the broad-line AGN with flat X-ray spectra
which are required to explain the CXRB.

\section{Critical Assumptions}
Currently, the latest
AGN-synthesis models of the CXRB spectrum and XLF
make a number of assumptions about the source spectra which may not
be altogether correct or justified. We examine the most
important ones below.

\subsection{Intrinsic AGN continuum: Hard X-ray Slope}

The intrinsic hard X-ray spectral photon
index of both the type 1 and type 2
AGN is universally assumed to be $\Gamma = 2$  
in AGN synthesis models.
While it is true that the mean photon index measured for 
samples of AGN peak at this value 
(e.g. \cite{nand99}), the dispersion is large.
This is most dramatically illustrated in Fig. 6 of \cite{geor99} 
which shows a range of photon indices 
($\Delta \Gamma > 1$; see Fig. 5 in \cite{geor99} ) found for a sample
of PG quasars. Aside from these examples, we know
there are a significant number of sources which do not 
have $\Gamma =2$ and may
have {\it intrinsically flat X-ray spectra}, perhaps too many to
ignore. 
This is not new of course: we have known for a long time that the
one of the brightest quasars, 3C 273, has $\Gamma \sim 1.5$.
Also, it appears that high-redshift quasars may generally have
flatter X-ray spectra \cite{vign99}.

The point is that we do not need to be constrained to assume
$\Gamma = 2$ for AGN synthesis models of the CXRB because we do not
know what value (or range) is relevant for the bulk of the sources
that make the CXRB. 

\subsection{Intrinsic AGN continuum: Soft Excess}

The latest AGN synthesis models still assume that a soft excess
above the basic $\Gamma = 2$ power law is a ubiquitous feature of
the intrinsic continuum in both type 1 and type 2 AGN (e.g.
\cite{gill99a}, \cite{coma99}).
Specifically, the models assume that the intrinsic power law steepens
to $\Gamma = 2.3$ below 1.5 keV.
Yet the latest studies of quasars show that this is simply not
borne out by observation. For example \cite{geor99} find
only 5 out of 14 PG quasars show a soft excess in the \asca band.
Also, \cite{reev97} 
find at most 2 out of 24 radio-loud and
radio-quiet quasars with a soft excess in the \asca band.
Sure, \rosat / \pspc studies all show that quasar spectra are
significantly steeper in the 0.1--2.4 keV band than in the 2--10 keV
band. Aside from systematic problems with the \pspc response 
(e.g. \cite{pspcal}, \cite{miya98})
there may well be a soft-excess component that is common {\it below 0.5}
keV. However, in the 0.5--2 keV band, which is relevant here, there
is no evidence whatsoever from \asca or \rosat
that a soft excess in the 0.5--2 keV band is ubiquitous.
The lack of evidence for an
intrinsic soft excess is even more severe in type 2 AGN, in which the 
intrinsic soft
X-ray spectrum cannot be seen in
most cases. It is either cut-off or is dominated by complex line-emission.

The ubiquity of a soft X-ray excess in AGN is a myth left over from
the days of low-energy resolution spectroscopy in the eighties,
before it was realised that photoionized absorption edges are
common in AGN and that these can mimic a steepening of the soft X-ray
spectrum with low-energy resolution data. However, the mimicking is
only approximate and the two types of spectra differ in important ways
(e.g. see {\cite{geor99}).

\subsection{Intrinsic AGN continuum: High-energy cutoff}

AGN synthesis models generally assume an exponential cutoff of
around 300 keV or so (e.g. \cite{coma99}, \cite{gill99a}). The observational situation is
that this is a very difficult parameter to measure with current
instrumentation. A good review can be found in \cite{matt98}.
It can be seen that the error bars are large and there is quite a
range in the allowed cutoff values. In fact, one of the best measurements
(for the bright AGN NGC 4151) is low at,  $\sim 70 \pm 15$ keV
(see Table 1 in \cite{matt98}). For the sources which make the bulk of the CXRB, we do
not know whether low values or high values are relevant because we
have no spectral information in that energy band for the weak
sources identified in deep surveys. The distribution of the high-energy
turnover in the sources that make the CXRB is important for reproducing
the spectral bump at $\sim 30$ keV.

\subsection{Absorption in Type 2 AGN}

Several attempts have been made recently to characterize the
absorbing column distribution in type 2 AGN (e.g. \cite{bass99}, 
\cite{maio98}, \cite{risa99}), to use as
input to AGN synthesis models of the CXRB. It has
been known that absorbing columns deduced using
data only  below
10 keV, not taking into account higher-energy data,
can be wrong. However, type 2 AGN have generally been too weak
to obtain good spectra above 10 keV. Only recently has it been
demonstrated, using \bsax data,
that the absorbing system parameters deduced previously for some
well-known sources have been completely wrong. For example,
the X-ray spectrum of Circinus galaxy was thought to be due to pure
Compton reflection. However, \bsax discovered a transmission
component at hard X-ray energies \cite{matt99}. 
Also, Mkn 3 was previously thought
to be Compton-thin but \bsax utilizing higher energy data
actually shows it to be Compton-thick \cite{capp99}.
Since the absorbing column distributions compiled so far (e.g. in 
\cite{bass99}, \cite{maio98}, \cite{risa99}) are
based on previous data usually without the benefit of hard X-ray
data, it has to be said that the true column density distribution is
actually unknown. In any case, the distributions that have been
compiled are only for local sources and not the ones that contribute
to the bulk of the CXRB.

Another uncertainty in AGN synthesis models is how to treat type 2
QSO. A few have been discovered (e.g. \cite{t2qsoa}, \cite{t2qsob}) but their
existence has also been disputed (e.g. \cite{halp99}).

\subsection{The ratio of Type 2/Type 1 AGN}
AGN synthesis models either assume a value for this ratio or predict it.
There are three problems with this ratio. The first is that it is based
on optical classification and it is explicitly assumed
in the models that type 1
AGN are not subject to X-ray absorption. The latter is simply not true
(see \S 2.6). Thus, {\it the model ratio and the observed ratio are two
completely different quantities}. The model ratio is  an {\it X-ray
ratio} but the observed ratio is an {\it optical ratio}.
The second problem is that reports of the observed ratio
cover an order of magnitude in range, from $\sim 1-10$
(e.g. see discussion in \cite{gill99a}).
The third problem is that the ratio may be completely different at high
redshifts.
Currently, the type 2/type 1 ratio as it
stands is not a very useful parameter.

\subsection{Absorption in Type 1 AGN}

%
\begin{figure}[t]
\vspace{-2cm}
\centerline{\vbox{
\psfig{figure=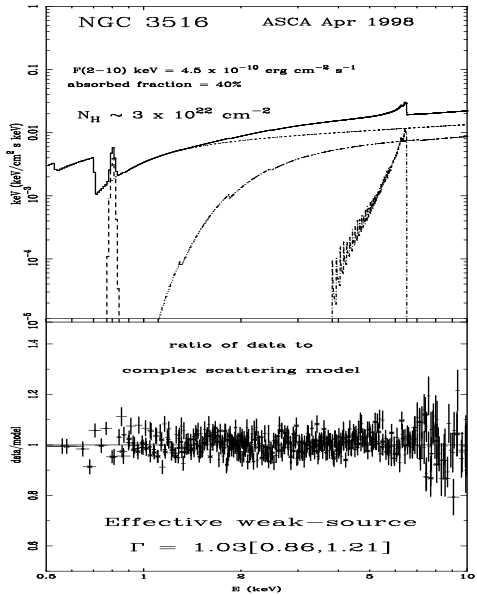,height=11cm}
}}
\vspace{-1cm}
\caption[]{ Best-fitting model ($\nu F_{\nu}$)
and residuals to the Seyfert 1 galaxy
NGC 3516 (see \S 2.6). The X-ray spectrum is very flat;
if this source were very faint it would be a simple power-law with
$\Gamma = 1.03$. }
\end{figure}

All AGN synthesis models ignore X-ray absorption in type 1 AGN.
Here we wish to demonstrate that even a small amount of absorption
can make the 0.5--10 keV X-ray spectrum flatter than the CXRB
spectrum. Complexity in the form of photoionized absorbers in
type 1 AGN is common (e.g. \cite{geor99}, \cite{nand99}) yet this is ignored in modeling
the CXRB. Consider NGC 3516 for example, which is a well-known broad-line
AGN. Fig. 1 shows the best-fitting 
absorption plus scattering model 
(including photoionization edge features) and residuals to \asca
data for NGC 3516. 
Such a model universally describes the main features
of the \asca spectra of many type 1 and type 2 AGN (e.g. \cite{serl96}).
The absorbing column is only $\sim 3 \times 10^{22} \rm \ cm^{-2}$.
While the intrinsic power law is steep ($\Gamma \sim 1.8$), the
overall spectrum is {\it extremely flat}. In fact if this source
was much fainter, say with a 2--10 keV flux of $10^{-13}
\rm \ ergs \ cm^{-2} \ s^{-1}$, a 100 Ks simulation shows that we
would measure a slope of $\Gamma = 1.03^{+0.18}_{-0.17}$
(assuming no extra absorption, which is what is done with very
faint sources detected in deep surveys).

NGC 3516 is not an isolated example. There are many other examples
in the literature of {\it broad-line AGN with overall flat
X-ray spectra}. NGC 3227, NGC 4151  are two other notable
examples. If these sources were very faint, simulations similar
to those for NGC 3516 show that we would measure $\Gamma = 0.40^{+0.20}
_{-0.25}$ and $\Gamma = -0.07^{+0.14}_{-0.28}$ respectively.
We also note that \cite{geor99} found a heavily absorbed quasar,
PG $1411+442$, in their
PG example with $N_{H} \sim 10^{23} \rm \ cm^{-2}$ which consequently
has a very flat X-ray spectrum.

\subsection{Scattering Fraction}

The value of the fraction of the observed continuum in type
2 AGN which is due to scattering and not subject
to the absorption suffered by the direct continuum, is typically
assumed to be $\sim 2-4\%$. Although authors of AGN synthesis
models admit there might be a small but real range in this parameter,
we suggest that the range is very large. For example we measure
$\sim 40\%$ for NGC 2992 and $0.2 \%$ for NGC 7172; these numbers
represent approximately
the extreme values we have measured for a large number
of objects.

\section{Alternative Models of the CXRB}

Other categories of objects have been proposed to explain some or
all of the CXRB. Notably starbursts 
(e.g. \cite{grif90}, \cite{mora99}, \cite{yaqo95}), LINERS (e.g. \cite{n1052}),
elliptical galaxies,  and naked ADAFs (\cite{dimat99a}, \cite{dimat99b}).
It remains to be seen whether starbursts can achieve a sizable
contribution to the CXRB. Only some LINER X-ray spectra
are flat but most are steep \cite{serl96}. Likewise, although elliptical
galaxies may have a hard X-ray tail 
\cite{dimat99a}, their overall spectrum
compared to the CXRB is too steep. As for naked ADAFs, none have been
recognizably observed so far and they require fine-tuning of the
accretion rate.

\section{Conclusions}

We conclude that the sources
having broad optical lines and flat X-ray spectra, which are
showing up in the deep surveys at the
faintest fluxes, need not be a new, mysterious population of
undiscovered types of X-ray source. The source spectral assumptions
which go into current AGN synthesis models have not caught up with
improved spectral measurements of AGN over the last decade.
For example, X-ray absorption in type 1 AGN is common and soft excesses
are not ubiquitous. Intrinsic type 1 X-ray spectra can also be flat. 
The flat X-ray spectrum type 1 AGN we already know of may simply be more
abundant at the faintest fluxes. 
It will be interesting to see what sources {\it Chandra} will show up
in deep hard surveys \cite{annh99}.  

{\bf Questions}
A couple of important questions were raised.

(1) {\it Isn't this what people are doing anyway? What have you done that
is different?}
While it is true that AGN synthesis models already include both 
absorbed and unabsorbed AGN, and it may be simply a matter of which
are called type 1 and type 2,
what people are doing is modeling
all type 1 AGN without X-ray absorption and then try to predict the
hard counts from the soft counts
(or vice-versa) using $\Gamma = 2$. They then
find a discrepancy which cannot easily be explained if one adheres
to incorrect assumptions about the source spectra (\S 2).
It has been
demonstrated that the discrepancy can be resolved if one empirically
assumes a flatter slope \cite{giom98}. 
This flatter slope for type 1 AGN may be real,
as has been discussed above. What people are also doing is to attempt to give
interpretation to the model type 2/type 1 ratio. One cannot take seriously 
interpretations of models of the CXRB spectrum and source
counts if the source spectra are
not modeled correctly.

(2) {\it Earlier we saw some spectral fitting results of faint sources
in deep surveys, but these had steep X-ray slopes, not flat}. \\
Those spectral slopes quoted were obtained by fitting a model with
intrinsic absorption included. This is not what is relevant for
comparison with the slope of the CXRB. One must fit the spectrum of
those sources without extra absorption if one is to compare with
the CXRB. The spectra are flat.

\vfill
\end{document}